\documentclass[aps,prb,twocolumn,epsfig,floats,superscriptaddress]{revtex4}
\usepackage{graphicx,epsf,natbib,amsmath,latexsym,amssymb,multirow}
\def\be{\begin{equation}}
\def\ee{\end{equation}}
\def\bea{\begin{eqnarray}}
\def\eea{\end{eqnarray}}
\usepackage{color}

\begin{document}
\title{Long-range Interaction Induced Phases in Weyl Semimetals}
\author{Huazhou Wei}
\affiliation{Department of Physics and Astronomy, University of California,
Riverside, CA 92521}
\author{Sung-Po Chao}
\affiliation{Physics Division, National Center for Theoretical Science, Hsinchu, 30013, Taiwan, R.O.C.}
\affiliation{Physics Department, National Tsing Hua University, Hsinchu, 30013, Taiwan, R.O.C.}
\author{Vivek Aji}
\affiliation{Department of Physics and Astronomy, University of California,
Riverside, CA 92521}

\date{\today}

\begin{abstract}
The interplay of spin orbit coupling and electron electron interaction condensing new phases of matter is an important new 
phenomena in solid state physics. In this paper we explore the nature of excitonic phases induced in Weyl semimetals by
long range Coulomb repulsion. Its has been previously shown that short range repulsion leads to a ferromagnetic insulator while short range attraction results in a 
charge density wave state. Here we show the the charge density wave is the energetically favored state in the presence of long range repulsion. 
\end{abstract}


\maketitle

\section{Introduction}
Over the last decade exciting new states of matter have been discovered in systems
with strong spin orbit coupling\cite{haldane, kane,zhangh,konig,teo, zhangrevmodphys,fu, xu, fu1, roy, jemoore}. Topological insulators are a striking example of this
phenomena. An important characteristic of this system is that it has two non degenerate bands
touching at an odd number of isolated points in the two dimensional Brillouin zone, with the
energy dispersing linearly in the vicinity of such points. Analogous properties in three dimensions
are rarer and as such accidental degeneracies are harder to realize. An important development 
was the theoretical prediction that a class of oxides (pyrochlore iridates) had the potential to
stabilize such nodes\cite{xwan}. Linearly dispersing massless fermions in three dimensions satisfy the Weyl equation 
and such phases of matter are the Weyl semimetals\cite{weyl}. Associated with this novel low energy
sector is a number of anomalous properties\cite{yran, abjaji, nn, kharzeev, hosur, sidp, zyuzin3, pallab}.

While the phase is yet to be discovered in the Iridates, a number of other proposals were made
to realize the phenomena. Balents and Burkov\cite{bb} showed that a heterostructure
made up of alternating layers of magnetically doped topological insulator and normal insulator had
Weyl fermions in its low energy sector. An alternate route is to find materials which have a Dirac
dispersion in three dimensions and break the spin degeneracy by breaking either time reversal or inversion. 
Angle resolved photoemission spectroscopy measurements on Na$_{3}$Bi\cite{zwangtop, zkl} and Cd$_{3}$As$_{2}$\cite{www, sb,neupane} have
provided the first evidence for the existence of massless Dirac fermions. The latter also breaks inversion 
and has the potential to be a Weyl semimetal (with nodes separated in momentum space), but the data lacks the resolution to verify the claim.

Weyl nodes are robust against translationally invariant interactions due to the conservation of chirality.
To illustrate this let us consider a system with the chemical potential at the node where the spins in the conduction (valence) band parallel (antiparallel) 
to the momentum. Only matrix element that flip chirality can mix the two. Since such potentials don't occur, the nodes are topologically
stable. The only way to open a gap is to mix two nodes with opposite chirality. In Weyl semimetals, these are separated
in momentum space and only short ranged interactions lead to gaps in the spectrum.

For local repulsive interactions, there are two classes of particle-hole or excitonic phases that are realized, corresponding to translationally
invariant states and charge density waves (CDWs) modulated at the nesting vector of the the modes\cite{hwei, zwang}. The former is further classified as either preserving or
breaking inversion symmetry. The energetically most favored state is the inversion preserving ferromagnetic state. On the either hand, for short range attraction
the CDW has the lowest energy. Interestingly enough, the CDW also wins over the possible superconducting states\cite{weisc} for chemical potential at the nodes. 

In this paper we consider the effect of long range repulsive (unscreened Coulomb) interactions and study the particle hole
instabilities. We show that the fall off of interaction as a function of momenta allows for the CDW phase to be accessible, which is strictly 
forbidden for short range interaction. We compare the decrease in energy of the two most competitive states that open a gap and find that
the CDW is the more favorable one. Thus for chemical potential at the node the most favorable states are : 1) ferromagnetic insulator for short range 
repulsion and 2) CDW for short range attraction or long range repulsion. For completeness we also note that a BCS superconductor is energetically
favored for long range attraction. A material realization of a Weyl semimetal will fall into one of these categories and the corresponding correlated state
will be condensed.

The paper is organized as follows. In section II the model of Weyl semimetals with two nodes is introduced. Section III deals with the projection of interaction
to the low energy bands and establishes the possible particle hole instabilities. In section IV we derive and solve the gap equations for instabilities induced by
Coulomb repulsion. We summarize our results and compare it with the renormalization group analysis done by Maciejko and Nandkishore\cite{mn} in the last section. 

\section{Model}

To explore the symmetries and energetic of the excitonic phases, we simplify to the
case of two Weyl nodes and the general density density interactions
with Coulomb form. For two Weyl nodes at
$\vec{K}_{1} = K_{0}\hat{x}$ (labeled R) and $-\vec{K}_{1}=
-K_{0}\hat{x}$ (labeled L) with chiralities $+1$ and $-1$
the Hamiltonian is
\begin{equation}\label{H0}
H_{0 \pm} = \pm \hbar v
\sum_{\vec{k}}\psi_{\vec{k}\alpha}^{\dagger}\vec{\sigma}_{\alpha\beta}\cdot\left(\vec{k}\mp
\vec{K}_{0}\right)\psi_{\vec{k}\beta}
\end{equation}
where $v$ is the Fermi velocity and $\vec{\sigma} = \{\sigma_{x},
\sigma_{y}, \sigma_{z}\}$ is a vector of Pauli matrices. The
dispersion at each node is $\epsilon_{\vec{q}}=\pm \hbar
v\left|\vec{q}\right|$ centered around $\pm \vec{K}_{0}$, with
$\vec{q} =\vec{k}\mp \vec{K}_{0}$. The conduction (valence) band at
the R node has its spin parallel (anti-parallel) to $\vec{q}$, while
the opposite is true at the L node. The general spin independent
particle particle interaction takes the form
\begin{eqnarray}\nonumber
V&=&\sum_{\sigma,\sigma'}\int d\vec{r} d\vec{r}'
V\left(\vec{r}-\vec{r}'\right)\psi^{\dagger}_{\sigma'}\left(\vec{r}'\right)
\psi_{\sigma'}\left(\vec{r}'\right)\psi^{\dagger}_{\sigma}\left(\vec{r}\right)
\psi_{\sigma}\left(\vec{r}\right)
\\ \label{V1}&=&\sum_{\sigma,\sigma'}\sum_{\vec{k},
\vec{k}',\vec{q}}V(\vec{q})
\psi^{\dagger}_{\vec{k'}+\vec{q},\sigma'}
\psi_{\vec{k'},\sigma'}\psi^{\dagger}_{\vec{k}-\vec{q},\sigma}\psi_{\vec{k},\sigma}
\end{eqnarray}
Here $V(\vec{q})=\frac{1}{\Omega}\int d\vec{r} V(\vec{r})$ with
$\Omega$ being the volume of the system.
For the moment we do not make any assumptions on the nature of
the interactions. Since the Weyl physics is the low energy
description of a more general theory, we enforce an upper cutoff
in the momentum integrals (up to $|\vec{q}|=\Lambda/\hbar v$
with a cutoff energy $\Lambda$) around each Weyl point. We use
mean field approach to study the various excitonic phases
associated with different type of particle hole interactions.\\

\begin{figure}
\includegraphics[width=0.8\columnwidth, clip]{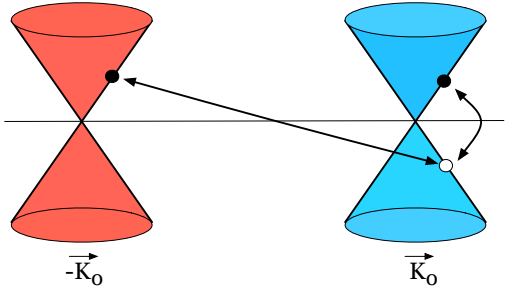}
\caption{The Weyl nodes at $\vec{K}_{0}$ and $-\vec{K}_{0}$ are shown. There are two types of particle hole instabilities
allowed. In the intranodal excitons lead to uniform order while the internodal  excitons lead to the Charge Density Waves at 
the nesting vector $\pm 2\vec{K}_{0}$.}\label{phases}
\end{figure}

There are two types of particle-hole excitations possible as shown in Fig.\ref{phases}: i) intra-nodal (occurring at zero
momentum) and ii) inter-nodal (occurring at a finite fixed momentum
associated with the nesting vector). These are excitonic
phases\cite{wk}, the former being the excitonic insulator (EI) while
the latter is the charge density wave (CDW). Unlike conventional
condensates in these sectors, the electron hole pairing of Weyl
fermions occurs both in even and odd parity channels. The latter are particle hole analogs of the superfluid 
phases of liquid $^{3}$He. In the next section we analyze the various possibilities by first projecting the interaction to low energy bands.

\section{Particle Hole instabilities}
To identify the symmetry breaking channels we project the electron electron interaction, written in spin basis in Eq.\ref{V1}, to the low energy chiral basis.  The corresponding wavefunctions are $c^{L,R}_{\vec{k},\pm}= \cos{\theta\over 2}\psi^{L,R}_{\vec{k}\uparrow}\pm \sin{\theta\over 2}e^{i\phi}\psi^{L,R}_{\vec{k}\downarrow}$, where $c^{L,R}_{\vec{k},\pm}$ is the annihilation operator for electrons in the band with chirality +/-  with momentum $\vec{k}$ measured from the L/R Weyl node.  The range of $\vec{k}$ is limited to lie within energy $2\Lambda$ around each node. Introducing these operators into Eq.\ref{V1} we get 16 terms of which only 6 satisfy momentum conservation in the allowed energy window.

\begin{figure}[h]
\includegraphics[width=0.5\columnwidth, clip]{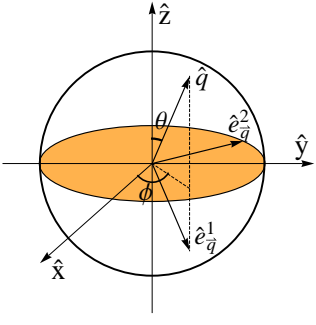}
\caption{The interaction shown in Eq.(\ref{fullV}) is a function of
three vectors ($\hat{q}$, $\hat{e}^{1}_{\vec{q}}$ and
$\hat{e}^{2}_{\vec{q}}$) that form a right handed coordinate system.
Each vector couples to an operator of distinct symmetry in the
particle hole channel.} \label{vectors}
\end{figure}

 To represent the interaction in a compact and physically transparent form we first define a coordinate system.
In the momentum space we choose the three orthogonal
unit vectors as follows: 1) $\hat{q} = \{\hat{q}_{x}, \hat{q}_{y},
\hat{q}_{z}\}$, the unit vector along $\vec{q}$;  2)
$\hat{e}_{\vec{q}}^{1} \equiv \hat{\theta}_{\vec{q}} =
\{\hat{q}_{x}\hat{q}_{z}/\sqrt{\hat{q}_{x}^{2}+
\hat{q}_{y}^{2}},\hat{q}_{y}\hat{q}_{z}/\sqrt{\hat{q}_{x}^{2}+\hat{q}_{y}^{2}},
-\sqrt{\hat{q}_{x}^{2}+\hat{q}_{y}^{2}}\}$; and 3)
$\hat{e}_{\vec{q}}^{2}\equiv \hat{\phi}_{\vec{q}} =
\{-\hat{q}_{y}/\sqrt{\hat{q}_{x}^{2}+\hat{q}_{y}^{2}},
\hat{q}_{x}/\sqrt{\hat{q}_{x}^{2}+\hat{q}_{y}^{2}},0\}$. By making
this choice $\hat{q}$, $\hat{e}_{\vec{q}}^{1} $ and
$\hat{e}_{\vec{q}}^{2} $ form a right handed coordinate system (see
Fig.\ref{vectors}). The unit sphere is spanned by the vector
$\hat{q}$ by two rotations. A prescription is to first pick a fixed axis perpendicular to $
\hat{e}_{\vec{q}}^{2}$. A rotation about  $
\hat{e}_{\vec{q}}^{2}$ followed by a rotation about the fixed axis traces out the unit sphere. For example if we choose the first to be
the $z$-axis, than the vector $\hat{e}^{2}_{\vec{q}}$, which is the
$\hat{\phi}$ in the spherical polar system, spans a unit circle
(perimeter of the shaded region in Fig.\ref{vectors}) and the vector
$\hat{e}^{1}_{\vec{q}}$, which is the corresponding $\hat{\theta}$,
spans the southern hemisphere. The following construction holds for
an arbitrary quantization axis $\hat{n}$, with the corresponding
polar and azimuthal angle for $\vec{q}$ defined in the coordinate
frame  $\{\hat{l}, \hat{m}, \hat{n}\}$. In the rest of the article
we use the $\{\hat{x}, \hat{y}, \hat{z}\}$ coordinate system.

Specializing to potentials related to particle hole interaction and
are even functions of $\vec{k}$, i.e. $V(\vec{k}) = V(-\vec{k})$,
the interaction in terms of $\hat{e}_{\vec{k}} =
\hat{e}_{\vec{k}}^{1}+\imath \hat{e}^{2}_{\vec{k}}$, is
\begin{widetext}
\begin{eqnarray}\nonumber
V=&-&\sum_{\vec{k},\vec{k}',n=\pm}\Bigg[
V(\vec{k}-\vec{k}')\frac{\hat{e}_{\vec{k}}\cdot
\hat{e}_{\vec{k}'}^{\ast}+ \hat{e}_{\vec{k}}
^{\ast}\cdot\hat{e}_{\vec{k'}}
}{4}\sum_{\tau=R,L}c_{\vec{k},n}^{\tau\dag
}c_{\vec{k},-n}^{\tau}c_{\vec{k}',-n}^{\tau\dag}c_{\vec{k}',n}^{\tau}
+V(\vec{k}-\vec{k'}-2\vec{K}_{0})\frac{\hat{e}_{\vec{k}}
\cdot\hat{e}_{\vec{k}'}+ \hat{e}_{\vec{k}}^{\ast}\cdot
\hat{e}_{\vec{k'}}^{\ast}}{2}\\
&\times&
c_{\vec{k},n}^{L\dag}c_{\vec{k},-n}^{L}c_{\vec{k}',-n}^{R\dag}c_{\vec{k}',n}^{R}
-\Big{[}2V(2\vec{K}_{0})-{V(\vec{k}-\vec{k}')}\Big{(}\hat{k}\cdot\hat{k}'+1\Big{)}\Big{]}
c_{\vec{k},n}^{L\dag}c_{\vec{k},-n}^{R}c_{\vec{k}',-n}^{R\dag}c_{\vec{k}',n}^{L}\Bigg]
\label{fullV}
\end{eqnarray}
\end{widetext}
The first and second term promote excitonic
instabilities with intra-nodal order parameter, i.e.
$\left<\sum_{\vec{k}}\vec{A}_{\vec{k}}c^{\tau\dag}_{\vec{k},n}c^{\tau}_{\vec{k},-n}\right>\neq
0$ with $\vec{A}_{\vec{k}}$ being even or odd function of $\vec{k}$. The
last term leads to inter-nodal CDW order with
$\left<\sum_{\vec{k}}\vec{A}_{\vec{k}}c^{\tau\dag}_{\vec{k},n}c^{\bar{\tau}}_{\vec{k},-n}\right>\neq
0$. Details of derivations for Eq.(\ref{fullV}) are shown in the Appendix \ref{ApA}.

For local interactions, $V(\vec{k})$ is a constant independent of momentum. In the CDW sector, the 
momentum independent couplings cancel out and the leading term is vectorial in nature and of the form $\vec{k}\cdot\vec{k}'$. Energetically
it is favorable to from a gapped state in the intra-nodal sector. For long range interaction $V(2\vec{K}_{0}) \ll V(\vec{k}-\vec{k}')$
as $|\vec{k}-\vec{k'}|\ll \Lambda/v\hbar \ll 2\vec{K}_{0}$. Thus the leading term in the CDW sector is the scalar $V(\vec{k}-{k}')$. We focus on the
two gapped states, the CDW sector and the excitonic insulator, to identify the energetically favorable state via mean field approach in the following section.

\section{Coulomb interaction}
The bare unscreened Coulomb interaction in momentum space is given by 
$V(\vec{q})=g/|\vec{q}|^2$, where $g$ is a positive coefficient. From the discussion above, 
 the dominant channels forming EI and CDW come
from the first and last two terms of Eq.(\ref{fullV}). Keeping
only the leading terms give:
\begin{equation}\label{fullV-leadingorder}
\begin{split}
V=&-\sum_{\vec{k},\vec{k}',n=\pm}\Bigg[V(\vec{k}-\vec{k}')c_{\vec{k},n}^{L\dag}c_{\vec{k},-n}^{R}c_{\vec{k}',-n}^{R\dag}c_{\vec{k}',n}^{L}\\
&+\frac{\cos(\phi-\phi')}{2}V(\vec{k}-\vec{k}')\sum_{\tau=R,L}c_{\vec{k},n}^{\tau\dag
}c_{\vec{k},-n}^{\tau}c_{\vec{k}',-n}^{\tau\dag}c_{\vec{k}',n}^{\tau}\Bigg]
\end{split}
\end{equation}
The first term is inter-nodal
interaction while the second term is intra-nodal interaction.
Within mean field analysis, the self-consistent equation for the particle-hole order parameter $\Delta_{a}(\vec{k'})=\langle \sum_n c_{\vec{k},n}^{\tau_a\dagger}c_{\vec{k},-n}^{\bar{\tau}_a}\rangle$ with Coulomb interaction is
\begin{equation}\label{DeltaEqu}
\Delta_{a}(\vec{k}) =
\sum_{\vec{k'}}V(\vec{k},\vec{k'}){{\Delta_{a}(\vec{k'})}\over{2E_{\vec{k'}}}}\tanh\frac{\beta
E_{\vec{k'}}}{2}
\end{equation}
Here $E_{\vec{k'}}=\sqrt{(\hbar
v|\vec{k'}|)^2+|\Delta_{a}(\vec{k'})|^2}$ and $a$ referring to either the EI ($\bar{\tau}_a=\tau_a$) or CDW phase ($\bar{\tau}_a=L$ and $\tau_a=R$, for example). The Coulomb interaction term in
momentum space is
$V(\vec{k},\vec{k'})=V(\vec{k}-\vec{k'})=g/|\vec{k}-\vec{k'}|^2$.  In the following two
subsections we establish the criteria for the formation of the CDW and EI phase within mean field.

\subsubsection{Inter-nodal part: CDW phase}

The angular dependence of the interaction is handled by performing an 
expansion in spherical harmonics and restricting our analysis to the leading terms\cite{ssg}. From Eq.(\ref{fullV-leadingorder}) the self consistent equation for the CDW phase is
\begin{equation}\label{DeltaEquCDW}
\Delta(\vec{k})=\frac{g}{2}
\sum_{\vec{k'}}\frac{1}{|\vec{k}-\vec{k'}|^2}\frac{\Delta(\vec{k'})}{\sqrt{(\hbar
vk')^2+|\Delta(\vec{k'})|^2}}
\end{equation}
We expand $\Delta(\vec{k})$ in spherical harmonics as
$\Delta(\vec{k})=\sum_{l=0}^{\infty}\sum_{m=-l}^{l}\Delta_{l}^{m}(|\vec{k}|)Y_l^m(\theta,\phi)$.
For the $|\vec{k}-\vec{k'}|^{-2}$ factor, we use twice of spherical harmonics expansion for $|\vec{k}-\vec{k'}|^{-1}$. The expression for $|\vec{k}-\vec{k'}|^{-1}$
is
\begin{equation}\label{kdiffexpansion}
\frac{1}{|\vec{k}-\vec{k'}|}=\sum_{l=0}^{\infty}
\sum_{m=-l}^l\frac{4\pi}{2l+1}\frac{k_<^l}{k_>^{l+1}}Y_l^{m^*}(\theta',\phi')Y_l^m(\theta,\phi)
\end{equation}
Here, $k_<$=min$(|\vec{k}|,|\vec{k'}|)$ and
$k_>$=max$(|\vec{k}|,|\vec{k'}|)$. The resulting gap equation to leading order (details of the calculation
are shown in the Appendix \ref{ApB}) is
\begin{equation}\label{afinal}
\begin{split}
\Delta_0^0(u)&=\frac{g}{4\pi^2\hbar v}
\int_0^u\Big{(}\frac{u'}{u}\Big{)}^2\frac{\Delta_0^0(u')du'}{\sqrt{u'^2+\frac{1}{4\pi}\Delta_0^0(u')^2}}\\
&+\frac{g}{4\pi^2\hbar v}\int_u^{\Lambda}\frac{\Delta_0^0(u')du'}{\sqrt{u'^2+\frac{1}{4\pi}\Delta_0^0(u')^2}}\\
\end{split}
\end{equation}
where $u=\hbar vk$. Taking derivatives
with respect to $u$ on both sides gives:
\begin{equation}
\frac{d\Delta_0^0(u)}{du}=\frac{-2}{u}\bigg{(}\Delta_0^0(u)-\frac{g}{4\pi^2\hbar
v}\int_u^{\Lambda}\frac{\Delta_0^0(u')du'}{\sqrt{u'^2+\frac{1}{4\pi}\Delta_0^0(u')^2}}\bigg{)}
\end{equation}
Hereafter we suppress the superscript and subscript of
$\Delta_0^0(u)$ for clarity. Multiplying $u$ on two sides and taking derivative
with respect to $u$ again gives:
\begin{equation}\label{afinaldifferential}
u\frac{d^2\Delta(u)}{du^2}+3\frac{d\Delta(u)}{du}+\frac{g}{2\pi^2\hbar
v}\frac{\Delta(u)}{\sqrt{u^2+\frac{1}{4\pi}\Delta(u)^2}}=0
\end{equation}
\begin{figure}
\includegraphics[width=0.75\linewidth]{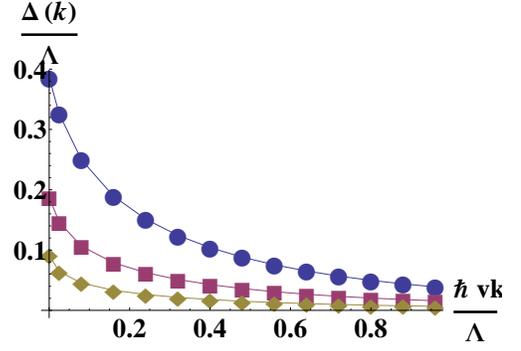}
\caption{Order parameter magnitude $\Delta_0^0(\vec{k})$ obtained by numerical iterations as a function of energy $\hbar v k$ for different interaction strengths $g/4\pi^2hv\sim 1.11$ (blue $\circ$), $g/4\pi^2hv\sim 0.93$ (purple $\Box$), and $g/4\pi^2hv\sim 0.74$ (brown $\diamond$) (color online).} \label{us0}
\end{figure}
\begin{figure}
\includegraphics[width=0.75\linewidth]{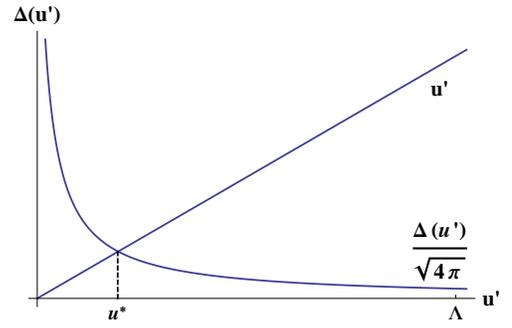}
\caption{ An approximate solution to Fig.\ref{us0} of the form $\Delta(u)\sim 1/u$ is plotted as well as a line with slop value equal to one. The point of intersection separates two regions. The contribution to the gap equation Eq.(\ref{afinal}) from $u<u^{\ast}$
is smaller than that of $u>u^{\ast}$ } \label{us}
\end{figure}
The solution for the order parameter $\Delta(u)$ as a function of energy $u$ shown in Fig.\ref{us0} is obtained numerically through numerical iterations. To evaluate Eq.(\ref{afinal}) analytically, we make the following observation: the dominant contribution to $\Delta(k)$
comes from the region of $u'\in[u^*,\Lambda]$, where $u^*$ is
determined by the condition $u^*=\Delta(u^*)/\sqrt{4\pi}$. In Fig.\ref{us} we plot the rough estimate for order parameter $\Delta(u)$ shown in Fig.\ref{us0} 
and identify the region that dominates the integrals in Eq.(\ref{afinal}). In the $u'>u^{\ast}$ region, $u'>\Delta(u')$. This condition allows us to
approximate Eq.(\ref{afinal}) and Eq.(\ref{afinaldifferential}) as
\begin{eqnarray}\nonumber
\Delta(u)&=&\frac{g}{4\pi^2\hbar v}\int_{u^*}^udu'\Big{(}\frac{u'}{u}\Big{)}^2\frac{\Delta(u')}{u'}\\\label{afinal-}
&+&\frac{g}{4\pi^2\hbar v}\int_u^{\Lambda}du'\frac{\Delta(u')}{u'}
\end{eqnarray}
and
\begin{equation}\label{afinaldifferential-}
u^2\frac{d^2\Delta(u)}{du^2}+3u\frac{d\Delta(u)}{du}+\frac{g}{2\pi^2\hbar
v}\Delta(u)=0
\end{equation}
respectively. General solution for Eq.(\ref{afinaldifferential-}) is:
\begin{equation}
\Delta(u)=Au^{-1-\sqrt{1-\frac{g}{2\pi^2\hbar
v}}}+Bu^{-1+\sqrt{1-\frac{g}{2\pi^2\hbar v}}}
\end{equation}
From Eq.(\ref{afinal-}), we have the boundary conditions for the
differential equation:
\begin{eqnarray}\nonumber
&&\left.\frac{d\Delta(u)}{du}\right|_{u\rightarrow u^*}=0\\
&&\left.\bigg{(}u\frac{d\Delta(u)}{du}+2\Delta(u)\bigg{)}\right|_{u\rightarrow\Lambda}=0
\end{eqnarray}
To satisfy these, we get a condition on $u^{\ast}$:
\begin{equation}
\frac{u^*}{\Lambda}=\bigg{(}\frac{1+\sqrt{1-\frac{g}{2\pi^2\hbar
v}}}{1-\sqrt{1-\frac{g}{2\pi^2\hbar
v}}}\bigg{)}^{\frac{1}{\sqrt{1-\frac{g}{2\pi^2\hbar v}}}}
\end{equation}
A valid solution occurs when the crossover scale is smaller than the upper cut off of the problem (i.e  $u^{\ast}/\Lambda <1$).
For small $g$ this condition is not satisfied. Thus a critical coupling is needed for the CDW phase to be stabilized. The condition  
required is  $1-\frac{g}{2\pi^2\hbar v}<0$. Defining $\sqrt{1-\frac{g}{2\pi^2\hbar v}}=i\alpha$ with
$\alpha=\sqrt{\frac{g}{2\pi^2\hbar v}-1}>0$, we have:
\begin{equation}
\frac{u^*}{\Lambda}=\bigg{(}\frac{1+i\alpha}{1-i\alpha}\bigg{)}^{\frac{1}{i\alpha}}
=e^{\frac{1}{i\alpha}\ln\frac{1+i\alpha}{1-i\alpha}}
\end{equation}
Since the modulus of $\frac{1+i\alpha}{1-i\alpha}$ is one, 
we denote $\frac{1+i\alpha}{1-i\alpha}=e^{i\varphi}$. This provides a solution of the form
\begin{equation}
\frac{u^*}{\Lambda}=e^{\frac{\varphi}{\alpha}}
\end{equation}
For $\frac{u^*}{\Lambda}<1$ we get
$\frac{\varphi}{\alpha}<0$, which leads to $\varphi<0$ as
$\alpha=\sqrt{\frac{g}{2\pi^2\hbar v}-1}$ is positive.
Furthermore $\cos\varphi=\frac{1-\alpha^2}{1+\alpha^2}$ and
$\sin\varphi=\frac{2\alpha}{1+\alpha^2}$ with $\alpha>0$ gives the condition
$\cos\varphi\in(-1,1)$ and $\sin\varphi\in(0,1)$. These are manifestly satisfied.
Thus the restriction on g comes from: $\frac{g}{2\pi^2\hbar v}-1>0$, yielding $g> 2\pi^2\hbar v$.\\


To summarize, for long range Coulomb interaction the inter nodal
instabilities require a minimum interaction strength for chemical
potential at the node. The minimum strength required is $2\pi^2\hbar v$.


\subsubsection{Intra-nodal part: EI phase}

The leading order coupling in the intra-nodal sector of Eq.(\ref{fullV-leadingorder}) is angle dependent and of the form:

\begin{equation}
\begin{split}
V_{intra}=&-\sum_{\vec{k},\vec{k'}}V(\vec{k}-\vec{k'})\frac{\cos(\phi-\phi')}{2}\times\\
&\sum_{\tau=R,L}\Big{(}c_{\vec{k}+}^{\tau\dagger}c_{\vec{k}-}^{\tau}c_{\vec{k'}-}^{\tau\dagger}c_{\vec{k'}+}^{\tau}
+c_{\vec{k}-}^{\tau\dagger}c_{\vec{k}+}^{\tau}c_{\vec{k'}+}^{\tau\dagger}c_{\vec{k'}-}^{\tau}\Big{)}\\
\end{split}
\end{equation}
Here we have expanded the summation on $n$. Within mean field, taking $\Delta(\vec{k})=\left\langle
c_{\vec{k}+}^{\tau\dagger}c_{\vec{k}-}^{\tau}\right\rangle$, the two terms are of the form $\Delta(\vec{k})\Delta(\vec{k'})^*+\Delta(\vec{k})^*\Delta(\vec{k'})$. They are equal
as is evident if we switch the dummy variables $\vec{k}$ and  $\vec{k'}$. The gap equation is
Eq.(\ref{DeltaEqu}) becomes:
\begin{equation}
\Delta(\vec{k})=\frac{g}{4}\sum_{\vec{k'}}\frac{e^{i\phi}e^{-i\phi'}
+e^{-i\phi}e^{i\phi'}}{|\vec{k}-\vec{k'}|^2}\frac{\Delta(\vec{k'})}{\sqrt{(\hbar
v|\vec{k'}|)^2+|\Delta(\vec{k'})|^2}}
\end{equation}
where we have written $\cos(\phi-\phi')$ explicitly. The solution takes the form
$\Delta(\vec{k})=\Delta(k)e^{i\phi}$, and the equation can be rewritten as 
\begin{equation}\label{DeltaEquEI}
\Delta(k)=\frac{g}{4}\sum_{\vec{k'}}\frac{1}{|\vec{k}-\vec{k'}|^2}\frac{\Delta(k')}{\sqrt{(\hbar
v|\vec{k'}|)^2+\Delta(k')^2}}
\end{equation}

Comparing Eq.(\ref{DeltaEquEI}) with Eq.(\ref{DeltaEquCDW}) which was derived for the CDW case, 
we find that the only difference is a factor of 2 arising from the angular dependent of the coupling. Therefore we can directly get the modified order
parameter equation and the corresponding differential equation:

\begin{equation}
\begin{split}
\Delta(u)=\frac{g}{8\pi^2\hbar
v}\bigg{[}&\int_0^{u}\Big{(}\frac{u'}{u}\Big{)}^2\frac{\Delta(u')du'}{\sqrt{u'^2+\Delta(u')^2}}\\
+&\int_u^{\Lambda}\frac{\Delta(u')du'}{\sqrt{u'^2+\Delta(u')^2}}\bigg{]}\\
\end{split}
\end{equation}
and
\begin{equation}
u\frac{d^2\Delta(u)}{du^2}+3\frac{d\Delta(u)}{du}+\frac{g}{4\pi^2\hbar
v}\frac{\Delta(u)}{\sqrt{u^2+\Delta(u)^2}}=0
\end{equation}
Following the same analysis as before we find a critical value of g given by $g>4\pi^2\hbar v$ for intra-nodal case.\\

 By comparing the two cases in
this long range Coulomb analysis we find that,
unlike the short range case\cite{hwei} where repulsion does not stabilize a gapped CDW phase, the possible ground state for unscreened Coulomb interaction
is the CDW phase.

\section{Conclusion}

The interplay of electron electron interaction and spin orbit coupling leads to a number of novel 
excitonic phases. Here we show that the range of the repulsive potential is crucial in determining 
the ground state. For short range interaction mean field studies show that a ferromagnetic excitonic 
insulator is energetically most favorable\cite{hwei}. If one considers the most general interaction allowed by symmetries,
beyond the Hubbard type, a renormalization group analysis finds the spin density wave (SDW) to be the leading symmetry breaking 
instabilty\cite{mn}. While the procedure is uncontrolled it is unbiased in that it does not a priori require a choice of order parameter.
It relies only on the generation of the minimum set of coupling constants, beyond the Hubbard repulsion, on projection to
the low energy sector. An interesting feature of the SDW, which it shares with the CDW, is that the electromagnetic response is rather
unusual in that it is axionic in character\cite{zwang, mn}. As such the physical realization of such modulated phases is of great interest. The CDW was shown to be stabilized by attractive short range interaction within mean field\cite{zwang}. In this paper we show that the CDW is also energetically favored for long range repulsive interaction. The fact that
the density of state is small and goes to zero at the nodes suggests that the screening in these materials is poor and that the interactions have finite
range. An unbiased renormalization group scheme on interactions with finite range could provide further support to this conjecture and a topic for future 
investigation.

\section{Acknowledgment}
H. Wei and V. Aji acknowledge the financial support by University of
California at Riverside through the initial complement. V. Aji also acknowledges the support of the Hellman foundation. S. P. Chao acknowledges the financial support by NCTS in Taiwan.

\appendix
\section{Derivation of Eq.(\ref{fullV})}\label{ApA}
The general spin independent particle particle interaction in momentum space is 
\begin{eqnarray}\nonumber\label{V}
V=\sum_{\sigma,\sigma'}\sum_{\vec{k},
\vec{k}',\vec{q}}V(\vec{q})
\psi^{\dagger}_{\vec{k'}+\vec{q},\sigma'}
\psi_{\vec{k'},\sigma'}\psi^{\dagger}_{\vec{k}-\vec{q},\sigma}\psi_{\vec{k},\sigma}
\end{eqnarray}
Here we denote $\vec{k}$, $\vec{k'}$ as incoming momenta and $\vec{k}-\vec{q}$, $\vec{k'}+\vec{q}$ as the outgoing momenta in the scattering process. Between two 
Weyl nodes (left ("$L$") and right ("$R$") nodes) there are total of $2^4=16$ possible scattering/interaction processes. Restricting to the low energy sector, conservation of energy and momentum leads to $6$ possible combinations for $(\vec{k},\vec{k'},\vec{k}-\vec{q},\vec{k'}+\vec{q})$: $(L,L,L,L)$, $(R,R,R,R)$, $(L,R,L,R)$, $(L,R,R,L)$, $(R,L,L,R)$, and $(R,L,R,L)$. These are
\begin{widetext}
\begin{eqnarray}\label{01}
&&V=\sum_{\vec{k},\vec{k'}}\sum_{\sigma,\sigma'}\Big[V(0)(\psi_{\vec{k'},\sigma'}^{L\dagger}\psi_{\vec{k'},\sigma'}^{L}+\psi_{\vec{k'},\sigma'}^{R\dagger}\psi_{\vec{k'},\sigma'}^{R})(\psi_{\vec{k},\sigma}^{L\dagger}\psi_{\vec{k},\sigma}^{L}+\psi_{\vec{k},\sigma}^{R\dagger}\psi_{\vec{k},\sigma}^{R})+V(\vec{k}-\vec{k'})(\psi_{\vec{k},\sigma'}^{L\dagger}\psi_{\vec{k'},\sigma'}^{L}\psi_{\vec{k'},\sigma}^{L\dagger}\psi_{\vec{k},\sigma}^{L}\\\nonumber
&&+\psi_{\vec{k},\sigma'}^{R\dagger}\psi_{\vec{k'},\sigma'}^{R}\psi_{\vec{k'},\sigma}^{R\dagger}\psi_{\vec{k},\sigma}^{R}+\psi_{\vec{k},\sigma'}^{R\dagger}\psi_{\vec{k'},\sigma'}^{L}\psi_{\vec{k'},\sigma}^{L\dagger}\psi_{\vec{k},\sigma}^{R}+\psi_{\vec{k},\sigma'}^{L\dagger}\psi_{\vec{k'},\sigma'}^{R}\psi_{\vec{k'},\sigma}^{R\dagger}\psi_{\vec{k},\sigma}^{L})+V(2\vec{K}_0)\psi_{\vec{k'}+2\vec{K}_0,\sigma'}^{R\dagger}\psi_{\vec{k'},\sigma'}^{L}\psi_{\vec{k}-2\vec{K}_0,\sigma}^{L\dagger}\psi_{\vec{k},\sigma}^{R}\\\nonumber
&&+V(-2\vec{K}_0)\psi_{\vec{k'}-2\vec{K}_0,\sigma'}^{L\dagger}\psi_{\vec{k'},\sigma'}^{R}\psi_{\vec{k}+2\vec{K}_0,\sigma}^{R\dagger}\psi_{\vec{k},\sigma}^{L}
+V(\vec{k}-\vec{k'}+2\vec{K}_0)\psi_{\vec{k}+2\vec{K}_0,\sigma'}^{R\dagger}\psi_{\vec{k'},\sigma'}^{R}\psi_{\vec{k'}-2\vec{K}_0,\sigma}^{L\dagger}\psi_{\vec{k},\sigma}^{L}+V(\vec{k}-\vec{k'}-2\vec{K}_0)\\\nonumber
&&\times \psi_{\vec{k}-2\vec{K}_0,\sigma'}^{L\dagger}\psi_{\vec{k'},\sigma'}^{L}\psi_{\vec{k'}+2\vec{K}_0,\sigma}^{R\dagger}\psi_{\vec{k},\sigma}^{R}\Big]
\end{eqnarray}
\end{widetext}
We express Eq.(\ref{01}) in the basis of unperturbed Hamiltonian $H_0$ in Eq.(\ref{H0}) which are obtained by solving the Sh\"{o}rdinger equations
\begin{eqnarray}\nonumber
&&\pm\hbar v\vec{q}\cdot\vec{\sigma} \chi_{\vec{q},\pm}^{R}=\pm \hbar v|\vec{q}|\chi_{\vec{q},\pm}^{R}\\\label{02}
&&\mp\hbar v\vec{q}\cdot\vec{\sigma} \chi_{\vec{q},\pm}^{L}=\pm \hbar v|\vec{q}|\chi_{\vec{q},\pm}^{L}
\end{eqnarray}
with $\vec{q}=\vec{k}\mp\vec{K}_0$ being the momenta relative to the  right/left node ("$\mp$" respectively). Defining
\begin{eqnarray}\label{02i}
\psi_{\vec{q},\pm}^{R(L)}=\chi_{\vec{q},\pm}^{R(L)}c_{\vec{q},\pm}^{R(L)}
\end{eqnarray}
 where $+ (-)$" labels the conduction(valence) band , and $c_{\vec{q},\pm}^{R(L)}$ is the corresponding fermion annihilation operator. The spin component of $\psi_{\vec{q},\pm,\sigma}^{R(L)}$ is   
\begin{eqnarray}\label{03}
\chi_{\vec{q},+}^{L}=\begin{pmatrix}
-\sin\frac{\theta}{2}e^{-i\phi} \\
\cos\frac{\theta}{2}
\end{pmatrix} ; 
\chi_{\vec{q},+}^{R}=\begin{pmatrix}
\cos\frac{\theta}{2}e^{-i\phi} \\
\sin\frac{\theta}{2}
\end{pmatrix}\\\nonumber
\chi_{\vec{q},-}^{L}=\begin{pmatrix}
\cos\frac{\theta}{2}e^{-i\phi} \\
\sin\frac{\theta}{2}
\end{pmatrix} ;
\chi_{\vec{q},-}^{R}=\begin{pmatrix}
-\sin\frac{\theta}{2}e^{-i\phi} \\
\cos\frac{\theta}{2}
\end{pmatrix}
\end{eqnarray}
The angles $\theta$ and $\phi$ are the polar and azimuthal angles of $\vec{q}$ (see fig.\ref{vectors}). In the particle-hole channel, $V(0)$ terms do not contribute due to conservation of chirality (i.e. $\chi_{\vec{q},+}^{R(L)\dagger}\chi_{\vec{q},-}^{R(L)}=\chi_{\vec{k}\mp\vec{K}_0,+}^{R(L)\dagger}\chi_{\vec{k}\mp\vec{K}_0,-}^{R(L)}=0$). For system with inversion symmetry we have $V(\vec{k})=V(-\vec{k})$, simplifying the interaction potential to the form
\begin{widetext}
\begin{eqnarray}\nonumber
V&=&\sum_{\vec{k},\vec{k'}}\sum_{\sigma,\sigma'}\Big[V(\vec{k}-\vec{k'})(\psi_{\vec{k},\sigma'}^{L\dagger}\psi_{\vec{k'},\sigma'}^{L}\psi_{\vec{k'},\sigma}^{L\dagger}\psi_{\vec{k},\sigma}^{L}+\psi_{\vec{k},\sigma'}^{R\dagger}\psi_{\vec{k'},\sigma'}^{R}\psi_{\vec{k'},\sigma}^{R\dagger}\psi_{\vec{k},\sigma}^{R}+2\psi_{\vec{k},\sigma'}^{L\dagger}\psi_{\vec{k'},\sigma'}^{R}\psi_{\vec{k'},\sigma}^{R\dagger}\psi_{\vec{k},\sigma}^{L})\\\label{04}
&+&2V(2\vec{K}_0)\psi_{\vec{k'}+2\vec{K}_0,\sigma'}^{R\dagger}\psi_{\vec{k'},\sigma'}^{L}\psi_{\vec{k}-2\vec{K}_0,\sigma}^{L\dagger}\psi_{\vec{k},\sigma}^{R}
+2V(\vec{k}-\vec{k'}-2\vec{K}_0)\psi_{\vec{k}-2\vec{K}_0,\sigma'}^{L\dagger}\psi_{\vec{k'},\sigma'}^{L}\psi_{\vec{k'}+2\vec{K}_0,\sigma}^{R\dagger}\psi_{\vec{k},\sigma}^{R}\Big]
\end{eqnarray}
\end{widetext}
To use the basis in Eq.(\ref{03}) we replace $\vec{k}$ (and $\vec{k'}$) by $\vec{k}=\vec{q}\pm\vec{K}_0$ for $\vec{k}$ at right/left node and then relabel 
the operators without specifying the origin. That is, we replace $\psi^{R/L}_{\vec{q}\pm\vec{K}_0,\sigma}$ by $\psi^{R/L}_{\vec{q},n,\sigma}$ as $R/L$ labels carry the same information and we specify the band index of the fermion operator by label $n=\pm$ (indicating conduction/valence band). We focus on the possible particle-hole pairing of the form $\psi_{\vec{q},\pm,\sigma}^{\alpha\dagger}\psi_{\vec{q},\mp,\bar{\sigma}}^{\beta}$. Note that spin orbit coupling implies that off diagonal pairing is allowed encoded by  $\sigma$ and $\bar{\sigma}$ not necessarily being the same. $\alpha$ and $\beta$ label left or right Weyl node. This yields
\begin{widetext}
\begin{eqnarray}\nonumber
V&=&-\sum_{\vec{q},\vec{q'}}\sum_{n_i=\pm}\sum_{\sigma,\sigma'}\Big[V(\vec{q}-\vec{q'})(\psi_{\vec{q},n_1,\sigma'}^{L\dagger}\psi_{\vec{q},n_2,\sigma}^{L}\psi_{\vec{q'},n_3,\sigma}^{L\dagger}\psi_{\vec{q'},n_4,\sigma'}^{L}+\psi_{\vec{q},n_1,\sigma'}^{R\dagger}\psi_{\vec{q},n_2,\sigma}^{R}\psi_{\vec{q'},n_3,\sigma}^{R\dagger}\psi_{\vec{q'},n_4,\sigma'}^{R})\\\label{05}&+&2V(\vec{q}-\vec{q'}-2\vec{K}_0)\psi_{\vec{q},n_1,\sigma'}^{L\dagger}\psi_{\vec{q},n_2,\sigma}^{L}\psi_{\vec{q'},n_3,\sigma}^{R\dagger}\psi_{\vec{q'},n_4,\sigma'}^{R}\\\nonumber
&-&2V(2\vec{K}_0)\psi_{\vec{q'},n_1,\sigma'}^{R\dagger}\psi_{\vec{q'},n_2,\sigma'}^{L}\psi_{\vec{q},n_3,\sigma}^{L\dagger}\psi_{\vec{q},n_4,\sigma}^{R}
+2V(\vec{q}-\vec{q'})\psi_{\vec{q},n_1,\sigma'}^{L\dagger}\psi_{\vec{q},n_2,\sigma}^{R}\psi_{\vec{q'},n_3,\sigma}^{R\dagger}\psi_{\vec{q'},n_4,\sigma'}^{L}\Big]
\end{eqnarray}
\end{widetext}
Eq.(\ref{05}) is structurally similar to Eq.(\ref{fullV}). The first three terms in  Eq.(\ref{05}) correspond to possible excitonic (particle-hole) pairings within the same Weyl node (or intra-nodal). The last two terms in  Eq.(\ref{05}) indicate possible pairings between two different nodes (inter-nodal). To get Eq.(\ref{fullV}) from Eq.(\ref{05}), we use the basis in Eq.(\ref{02i},\ref{03}) with creation and annihilation fermion operators chosen at different bands (one at conduction and another at valence band for the same momentum $\vec{q}$) and sum over the spin indices in Eq.(\ref{05}). In the intra-nodal part the first term:
\begin{widetext}
\begin{eqnarray}\nonumber
&&\sum_{n_i=\pm}\sum_{\sigma,\sigma'}\psi_{\vec{q},n_1,\sigma'}^{L\dagger}\psi_{\vec{q},n_2,\sigma}^{L}\psi_{\vec{q'},n_3,\sigma}^{L\dagger}\psi_{\vec{q'},n_4,\sigma'}^{L}\simeq
\sum_{\sigma,\sigma'}\Big(\psi_{\vec{q},+,\sigma'}^{L\dagger}\psi_{\vec{q},-,\sigma}^{L}\psi_{\vec{q'},-,\sigma}^{L\dagger}\psi_{\vec{q'},+,\sigma'}^{L}+\psi_{\vec{q},-,\sigma'}^{L\dagger}\psi_{\vec{q},+,\sigma}^{L}\psi_{\vec{q'},+,\sigma}^{L\dagger}\psi_{\vec{q'},-,\sigma'}^{L}\Big)\\\label{07}
&&=-\sum_{\sigma,\sigma'}\Big(\psi_{\vec{q},+,\sigma'}^{L\dagger}\psi_{\vec{q'},+,\sigma'}^{L}\psi_{\vec{q'},-,\sigma}^{L\dagger}\psi_{\vec{q},-,\sigma}^{L}+\psi_{\vec{q},-,\sigma'}^{L\dagger}\psi_{\vec{q'},-,\sigma'}^{L}\psi_{\vec{q'},+,\sigma}^{L\dagger}\psi_{\vec{q},+,\sigma}^{L}\Big)\\\nonumber
&&=-\left(c_{\vec{q},+}^{L\dagger}\chi_{\vec{q},+}^{L\dagger}\chi_{\vec{q'},+}^{L}c_{\vec{q'},+}^{L}c_{\vec{q'},-}^{L\dagger}\chi_{\vec{q'},-}^{L\dagger}\chi_{\vec{q},-}^{L}c_{\vec{q},-}^{L}+c_{\vec{q},-}^{L\dagger}\chi_{\vec{q},-}^{L\dagger}\chi_{\vec{q'},-}^{L}c_{\vec{q'},-}^{L}c_{\vec{q'},+}^{L\dagger}\chi_{\vec{q'},+}^{L\dagger}\chi_{\vec{q},+}^{L}c_{\vec{q},+}^{L}\right)\\\nonumber
&&=-\Bigg[c_{\vec{q},+}^{L\dagger}c_{\vec{q'},+}^{L}c_{\vec{q'},-}^{L\dagger}c_{\vec{q},-}^{L}\left(\sin\frac{\theta}{2}\sin\frac{\theta'}{2}e^{i(\phi-\phi')}+\cos\frac{\theta}{2}\cos\frac{\theta'}{2}\right)\left(\cos\frac{\theta}{2}\cos\frac{\theta'}{2}e^{i(\phi'-\phi)}+\sin\frac{\theta}{2}\sin\frac{\theta'}{2}\right)\\\nonumber
&&+c_{\vec{q},-}^{L\dagger}c_{\vec{q'},-}^{L}c_{\vec{q'},+}^{L\dagger}c_{\vec{q},+}^{L}\left(\cos\frac{\theta}{2}\cos\frac{\theta'}{2}e^{i(\phi-\phi')}+\sin\frac{\theta}{2}\sin\frac{\theta'}{2}\right)\left(\sin\frac{\theta}{2}\sin\frac{\theta'}{2}e^{i(\phi'-\phi)}+\cos\frac{\theta}{2}\cos\frac{\theta'}{2}\right)\Bigg]\\\nonumber
&&=c_{\vec{q},+}^{L\dagger}c_{\vec{q},-}^{L}c_{\vec{q'},-}^{L\dagger}c_{\vec{q'},+}^{L}\left(\frac{\sin\theta\sin\theta'}{2}+\frac{1+\cos\theta\cos\theta'}{2}\cos(\phi-\phi')-i\frac{\cos\theta+\cos\theta'}{2}\sin(\phi-\phi')\right)\\\nonumber
&&+c_{\vec{q},-}^{L\dagger}c_{\vec{q},+}^{L}c_{\vec{q'},+}^{L\dagger}c_{\vec{q'},-}^{L}\left(\frac{\sin\theta\sin\theta'}{2}+\frac{1+\cos\theta\cos\theta'}{2}\cos(\phi-\phi')+i\frac{\cos\theta+\cos\theta'}{2}\sin(\phi-\phi')\right)
\end{eqnarray}
\end{widetext}
The approximation made in the first line of Eq.(\ref{07}) is that the energy, as computed from the unperturbed Hamiltonian, is conserved. The summation over spin indices is carried out in the inner product of the two component spinor $\chi_{\vec{q},\pm}^{R/L}$. Similarly the second term in the intra-nodal part:
\begin{widetext}
\begin{eqnarray}\nonumber
&&\sum_{n_i=\pm}\sum_{\sigma,\sigma'}\psi_{\vec{q},n_1,\sigma'}^{R\dagger}\psi_{\vec{q},n_2,\sigma}^{R}\psi_{\vec{q'},n_3,\sigma}^{R\dagger}\psi_{\vec{q'},n_4,\sigma'}^{R}\simeq
\sum_{\sigma,\sigma'}\Big(\psi_{\vec{q},+,\sigma'}^{R\dagger}\psi_{\vec{q},-,\sigma}^{R}\psi_{\vec{q'},-,\sigma}^{R\dagger}\psi_{\vec{q'},+,\sigma'}^{R}+\psi_{\vec{q},-,\sigma'}^{R\dagger}\psi_{\vec{q},+,\sigma}^{R}\psi_{\vec{q'},+,\sigma}^{R\dagger}\psi_{\vec{q'},-,\sigma'}^{R}\Big)\\\label{08}
&&=-\sum_{\sigma,\sigma'}\Big(\psi_{\vec{q},+,\sigma'}^{R\dagger}\psi_{\vec{q'},+,\sigma'}^{R}\psi_{\vec{q'},-,\sigma}^{R\dagger}\psi_{\vec{q},-,\sigma}^{R}+\psi_{\vec{q},-,\sigma'}^{R\dagger}\psi_{\vec{q'},-,\sigma'}^{R}\psi_{\vec{q'},+,\sigma}^{R\dagger}\psi_{\vec{q},+,\sigma}^{R}\Big)\\\nonumber
&&=-\left(c_{\vec{q},+}^{R\dagger}\chi_{\vec{q},+}^{R\dagger}\chi_{\vec{q'},+}^{R}c_{\vec{q'},+}^{R}c_{\vec{q'},-}^{R\dagger}\chi_{\vec{q'},-}^{R\dagger}\chi_{\vec{q},-}^{R}c_{\vec{q},-}^{R}+c_{\vec{q},-}^{R\dagger}\chi_{\vec{q},-}^{R\dagger}\chi_{\vec{q'},-}^{R}c_{\vec{q'},-}^{R}c_{\vec{q'},+}^{R\dagger}\chi_{\vec{q'},+}^{R\dagger}\chi_{\vec{q},+}^{R}c_{\vec{q},+}^{R}\right)\\\nonumber
&&=-\Bigg[c_{\vec{q},+}^{R\dagger}c_{\vec{q'},+}^{R}c_{\vec{q'},-}^{R\dagger}c_{\vec{q},-}^{R}\left(\cos\frac{\theta}{2}\cos\frac{\theta'}{2}e^{i(\phi-\phi')}+\sin\frac{\theta}{2}\sin\frac{\theta'}{2}\right)\left(\sin\frac{\theta}{2}\sin\frac{\theta'}{2}e^{i(\phi'-\phi)}+\cos\frac{\theta}{2}\cos\frac{\theta'}{2}\right)\\\nonumber
&&+c_{\vec{q},-}^{R\dagger}c_{\vec{q'},-}^{R}c_{\vec{q'},+}^{R\dagger}c_{\vec{q},+}^{R}\left(\sin\frac{\theta}{2}\sin\frac{\theta'}{2}e^{i(\phi'-\phi)}+\cos\frac{\theta}{2}\cos\frac{\theta'}{2}\right)\left(\cos\frac{\theta}{2}\cos\frac{\theta'}{2}e^{i(\phi-\phi')}+\sin\frac{\theta}{2}\sin\frac{\theta'}{2}\right)\Bigg]\\\nonumber
&&=c_{\vec{q},+}^{R\dagger}c_{\vec{q},-}^{R}c_{\vec{q'},-}^{R\dagger}c_{\vec{q'},+}^{R}\left(\frac{\sin\theta\sin\theta'}{2}+\frac{1+\cos\theta\cos\theta'}{2}\cos(\phi-\phi')+i\frac{\cos\theta+\cos\theta'}{2}\sin(\phi-\phi')\right)\\\nonumber
&&+c_{\vec{q},-}^{R\dagger}c_{\vec{q},+}^{R}c_{\vec{q'},+}^{R\dagger}c_{\vec{q'},-}^{R}\left(\frac{\sin\theta\sin\theta'}{2}+\frac{1+\cos\theta\cos\theta'}{2}\cos(\phi-\phi')-i\frac{\cos\theta+\cos\theta'}{2}\sin(\phi-\phi')\right)
\end{eqnarray}
\end{widetext}
The third intra-nodal term (with $2V(\vec{q}-\vec{q'}-2\vec{K}_0)$ in front):
\begin{widetext}
\begin{eqnarray}\nonumber
&&\sum_{n_i=\pm}\sum_{\sigma,\sigma'}\psi_{\vec{q},n_1,\sigma'}^{L\dagger}\psi_{\vec{q},n_2,\sigma}^{L}\psi_{\vec{q'},n_3,\sigma}^{R\dagger}\psi_{\vec{q'},n_4,\sigma'}^{R}\simeq
\sum_{\sigma,\sigma'}\Big(\psi_{\vec{q},+,\sigma'}^{L\dagger}\psi_{\vec{q},-,\sigma}^{L}\psi_{\vec{q'},-,\sigma}^{R\dagger}\psi_{\vec{q'},+,\sigma'}^{R}+\psi_{\vec{q},-,\sigma'}^{L\dagger}\psi_{\vec{q},+,\sigma}^{L}\psi_{\vec{q'},+,\sigma}^{R\dagger}\psi_{\vec{q'},-,\sigma'}^{R}\Big)\\\label{09}
&&=-\sum_{\sigma,\sigma'}\Big(\psi_{\vec{q},+,\sigma'}^{L\dagger}\psi_{\vec{q'},+,\sigma'}^{R}\psi_{\vec{q'},-,\sigma}^{R\dagger}\psi_{\vec{q},-,\sigma}^{L}+\psi_{\vec{q},-,\sigma'}^{L\dagger}\psi_{\vec{q'},-,\sigma'}^{R}\psi_{\vec{q'},+,\sigma}^{R\dagger}\psi_{\vec{q},+,\sigma}^{L}\Big)\\\nonumber
&&=-\left(c_{\vec{q},+}^{L\dagger}\chi_{\vec{q},+}^{L\dagger}\chi_{\vec{q'},+}^{R}c_{\vec{q'},+}^{R}c_{\vec{q'},-}^{R\dagger}\chi_{\vec{q'},-}^{R\dagger}\chi_{\vec{q},-}^{L}c_{\vec{q},-}^{L}+c_{\vec{q},-}^{L\dagger}\chi_{\vec{q},-}^{L\dagger}\chi_{\vec{q'},-}^{R}c_{\vec{q'},-}^{R}c_{\vec{q'},+}^{R\dagger}\chi_{\vec{q'},+}^{R\dagger}\chi_{\vec{q},+}^{L}c_{\vec{q},+}^{L}\right)\\\nonumber
&&=-\Bigg[c_{\vec{q},+}^{L\dagger}c_{\vec{q'},+}^{R}c_{\vec{q'},-}^{R\dagger}c_{\vec{q},-}^{L}\left(-\sin\frac{\theta}{2}\cos\frac{\theta'}{2}e^{i(\phi-\phi')}+\sin\frac{\theta'}{2}\cos\frac{\theta}{2}\right)\left(-\sin\frac{\theta'}{2}\cos\frac{\theta}{2}e^{i(\phi'-\phi)}+\sin\frac{\theta}{2}\cos\frac{\theta'}{2}\right)\\\nonumber
&&+c_{\vec{q},-}^{L\dagger}c_{\vec{q'},-}^{R}c_{\vec{q'},+}^{R\dagger}c_{\vec{q},+}^{L}\left(-\sin\frac{\theta'}{2}\cos\frac{\theta}{2}e^{i(\phi-\phi')}+\sin\frac{\theta}{2}\cos\frac{\theta'}{2}\right)\left(-\sin\frac{\theta}{2}\cos\frac{\theta'}{2}e^{i(\phi'-\phi)}+\sin\frac{\theta'}{2}\cos\frac{\theta}{2}\right)\Bigg]\\\nonumber
&&=c_{\vec{q},+}^{L\dagger}c_{\vec{q},-}^{L}c_{\vec{q'},-}^{R\dagger}c_{\vec{q'},+}^{R}\left(\frac{\sin\theta\sin\theta'}{2}-\frac{1-\cos\theta\cos\theta'}{2}\cos(\phi-\phi')+i\frac{\cos\theta-\cos\theta'}{2}\sin(\phi-\phi')\right)\\\nonumber
&&+c_{\vec{q},-}^{L\dagger}c_{\vec{q},+}^{L}c_{\vec{q'},+}^{R\dagger}c_{\vec{q'},-}^{R}\left(\frac{\sin\theta\sin\theta'}{2}-\frac{1-\cos\theta\cos\theta'}{2}\cos(\phi-\phi')-i\frac{\cos\theta-\cos\theta'}{2}\sin(\phi-\phi')\right)
\end{eqnarray}
\end{widetext}
The first inter-nodal term (with $2V(2\vec{K}_0)$ in front):
\begin{widetext}
\begin{eqnarray}\nonumber
&&\sum_{n_i=\pm}\sum_{\sigma,\sigma'}\psi_{\vec{q'},n_1,\sigma'}^{R\dagger}\psi_{\vec{q'},n_2,\sigma'}^{L}\psi_{\vec{q},n_3,\sigma}^{L\dagger}\psi_{\vec{q},n_4,\sigma}^{R}\simeq
\sum_{\sigma,\sigma'}\Big(\psi_{\vec{q'},+,\sigma'}^{R\dagger}\psi_{\vec{q'},-,\sigma'}^{L}\psi_{\vec{q},-,\sigma}^{L\dagger}\psi_{\vec{q},+,\sigma}^{R}+\psi_{\vec{q'},-,\sigma'}^{R\dagger}\psi_{\vec{q'},+,\sigma'}^{L}\psi_{\vec{q},+,\sigma}^{L\dagger}\psi_{\vec{q},-,\sigma}^{R}\Big)\\\nonumber
&&=\left(c_{\vec{q'},+}^{R\dagger}\chi_{\vec{q'},+}^{R\dagger}\chi_{\vec{q'},-}^{L}c_{\vec{q'},-}^{L}c_{\vec{q},-}^{R\dagger}\chi_{\vec{q},-}^{L\dagger}\chi_{\vec{q},+}^{R}c_{\vec{q},+}^{R}+c_{\vec{q'},-}^{R\dagger}\chi_{\vec{q'},-}^{R\dagger}\chi_{\vec{q'},+}^{L}c_{\vec{q'},+}^{L}c_{\vec{q},+}^{R\dagger}\chi_{\vec{q},+}^{L\dagger}\chi_{\vec{q},-}^{R}c_{\vec{q},-}^{R}\right)\\\label{10}
&&=\left(c_{\vec{q'},+}^{R\dagger}c_{\vec{q'},-}^{L}c_{\vec{q},-}^{R\dagger}c_{\vec{q},+}^{R}+c_{\vec{q'},-}^{R\dagger}c_{\vec{q'},+}^{L}c_{\vec{q},+}^{R\dagger}c_{\vec{q},-}^{R}\right)
\end{eqnarray}
\end{widetext}
The second inter-nodal term (with $2V(\vec{q}-\vec{q'})$ in front):
\begin{widetext}
\begin{eqnarray}\nonumber
&&\sum_{n_i=\pm}\sum_{\sigma,\sigma'}\psi_{\vec{q},n_1,\sigma'}^{L\dagger}\psi_{\vec{q},n_2,\sigma}^{R}\psi_{\vec{q'},n_3,\sigma}^{R\dagger}\psi_{\vec{q'},n_4,\sigma'}^{L}\simeq\sum_{\sigma,\sigma'}\left(\psi_{\vec{q},+,\sigma'}^{L\dagger}\psi_{\vec{q},-,\sigma}^{R}\psi_{\vec{q'},-,\sigma}^{R\dagger}\psi_{\vec{q'},+,\sigma'}^{L}+\psi_{\vec{q},-,\sigma'}^{L\dagger}\psi_{\vec{q},+,\sigma}^{R}\psi_{\vec{q'},+,\sigma}^{R\dagger}\psi_{\vec{q'},-,\sigma'}^{L}\right)\\\label{11}
&&=-\sum_{\sigma,\sigma'}\left(\psi_{\vec{q},+,\sigma'}^{L\dagger}\psi_{\vec{q'},+,\sigma'}^{L}\psi_{\vec{q'},-,\sigma}^{R\dagger}\psi_{\vec{q},-,\sigma}^{R}+\psi_{\vec{q},-,\sigma'}^{L\dagger}\psi_{\vec{q'},-,\sigma'}^{L}\psi_{\vec{q'},+,\sigma}^{R\dagger}\psi_{\vec{q},+,\sigma}^{R}\right)\\\nonumber
&&=-\left(c_{\vec{q},+}^{L\dagger}\chi_{\vec{q},+}^{L\dagger}\chi_{\vec{q'},+}^{L}c_{\vec{q'},+}^{L}c_{\vec{q'},-}^{R\dagger}\chi_{\vec{q'},-}^{R\dagger}\chi_{\vec{q},-}^{R}c_{\vec{q},-}^{R}+c_{\vec{q},-}^{L\dagger}\chi_{\vec{q},-}^{L\dagger}\chi_{\vec{q'},-}^{L}c_{\vec{q'},-}^{L}c_{\vec{q'},+}^{R\dagger}\chi_{\vec{q'},+}^{R\dagger}\chi_{\vec{q},+}^{R}c_{\vec{q},+}^{R}\right)\\\nonumber
&&=-\Bigg[c_{\vec{q},+}^{L\dagger}c_{\vec{q'},+}^{L}c_{\vec{q'},-}^{R\dagger}c_{\vec{q},-}^{R}\left(\sin\frac{\theta}{2}\sin\frac{\theta'}{2}e^{i(\phi-\phi')}+\cos\frac{\theta}{2}\cos\frac{\theta'}{2}\right)\left(\sin\frac{\theta}{2}\sin\frac{\theta'}{2}e^{i(\phi'-\phi)}+\cos\frac{\theta}{2}\cos\frac{\theta'}{2}\right)\\\nonumber
&&+c_{\vec{q},-}^{L\dagger}c_{\vec{q'},-}^{L}c_{\vec{q'},+}^{R\dagger}c_{\vec{q},+}^{R}\left(\cos\frac{\theta}{2}\cos\frac{\theta'}{2}e^{i(\phi-\phi')}+\sin\frac{\theta}{2}\sin\frac{\theta'}{2}\right)\left(\cos\frac{\theta}{2}\cos\frac{\theta'}{2}e^{i(\phi'-\phi)}+\sin\frac{\theta}{2}\sin\frac{\theta'}{2}\right)\Bigg]\\\nonumber
&&=\left(c_{\vec{q},+}^{L\dagger}c_{\vec{q},-}^{R}c_{\vec{q'},-}^{R\dagger}c_{\vec{q'},+}^{L}+c_{\vec{q},-}^{L\dagger}c_{\vec{q},+}^{R}c_{\vec{q'},+}^{R\dagger}c_{\vec{q'},-}^{L}\right)\left(\frac{1+\cos\theta\cos\theta'}{2}+\frac{\sin\theta\sin\theta'}{2}\cos(\phi-\phi')\right)
\end{eqnarray}
\end{widetext}
The final step to reach Eq.(\ref{fullV}) is to replace the trigonometry expressions in Eq.(\ref{07})-Eq.(\ref{11}) by the unit vectors $(\hat{q},\hat{e}_{\vec{q}}^1,\hat{e}_{\vec{q}}^2)$ defined in the Fig.\ref{vectors}. With $\hat{e}_{\vec{q}}=\hat{e}_{\vec{q}}^1+i\hat{e}_{\vec{q}}^2$ we have
\begin{widetext}
\begin{eqnarray}\label{a1}
&&\frac{(\hat{e}_{\vec{q}}\cdot \hat{e}_{\vec{q'}}^{\ast}+\hat{e}_{\vec{q}}^{\ast}\cdot \hat{e}_{\vec{q'}})}{4}=\frac{\sin\theta\sin\theta'}{2}+\frac{1+\cos\theta\cos\theta'}{2}\cos(\phi-\phi')\\
&&\frac{(\hat{q}+\hat{q'})\cdot(\hat{e}_{\vec{q}}^2\times \hat{e}_{\vec{q'}}^2)}{-2}=\frac{(\cos\theta+\cos\theta')}{2}\sin(\phi-\phi')\\
&&\frac{(\hat{e}_{\vec{q}}\cdot \hat{e}_{\vec{q'}}+\hat{e}_{\vec{q}}^{\ast}\cdot \hat{e}_{\vec{q'}}^{\ast})}{4}=\frac{\sin\theta\sin\theta'}{2}-\frac{1-\cos\theta\cos\theta'}{2}\cos(\phi-\phi')\\
&&\frac{(\hat{q}-\hat{q'})\cdot(\hat{e}_{\vec{q}}^2\times \hat{e}_{\vec{q'}}^2)}{-2}=\frac{(\cos\theta-\cos\theta')}{2}\sin(\phi-\phi')\\\label{a2}
&&\frac{(\hat{q}\cdot \hat{q'}+1)}{2}=\frac{1+\cos\theta\cos\theta'}{2}+\frac{\sin\theta\sin\theta'}{2}\cos(\phi-\phi')
\end{eqnarray}
\end{widetext}
Using Eq.(\ref{a1})-Eq.(\ref{a2}) in Eq.(\ref{07})-Eq.(\ref{11}) and summing (or relabeling) over $\vec{q}$ and $\vec{q'}$ we get Eq.(\ref{fullV})
\begin{widetext}
\begin{eqnarray}\nonumber
V=&-&\sum_{\vec{q},\vec{q'},n=\pm}\Bigg[
V(\vec{q}-\vec{q'})\frac{\hat{e}_{\vec{q}}\cdot
\hat{e}_{\vec{q'}}^{\ast}+ \hat{e}_{\vec{q}}
^{\ast}\cdot\hat{e}_{\vec{q'}}
}{4}\sum_{\tau=R,L}c_{\vec{q},n}^{\tau\dag
}c_{\vec{q},-n}^{\tau}c_{\vec{q'},-n}^{\tau\dag}c_{\vec{q'},n}^{\tau}
+V(\vec{q}-\vec{q'}-2\vec{K}_{0})\frac{\hat{e}_{\vec{q}}
\cdot\hat{e}_{\vec{q'}}+ \hat{e}_{\vec{q}}^{\ast}\cdot
\hat{e}_{\vec{q'}}^{\ast}}{2}\\
&\times&
c_{\vec{q},n}^{L\dag}c_{\vec{q},-n}^{L}c_{\vec{q'},-n}^{R\dag}c_{\vec{q'},n}^{R}
-\Big{[}2V(2\vec{K}_{0})-{V(\vec{q}-\vec{q}')}\Big{(}\hat{q}\cdot\hat{q'}+1\Big{)}\Big{]}
c_{\vec{q},n}^{L\dag}c_{\vec{q},-n}^{R}c_{\vec{q'},-n}^{R\dag}c_{\vec{q'},n}^{L}\Bigg]
\end{eqnarray}
\end{widetext}

\section{Derication of the Gap Equation}\label{ApB}

\begin{widetext}
Expanding $\Delta(\vec{k})$ and $1/|\vec{k}-\vec{k'}|$
in spherical harmonics, the self-consistent
equation for excitonic order parameter is

\begin{equation}\label{a1}
\begin{split}
&\sum_{l_3m_3}\Delta_{l_3}^{m_3}(k)Y_{l_3}^{{m_3}^*}(\theta,\phi)=
\frac{g}{16\pi^3}\int k'^2\,dk'\int_0^{\pi} \sin\theta'\,d\theta'\int_0^{2\pi}\,d\phi'\times\\
&\sum_{l_1m_1}\sum_{l_2m_2}\frac{4\pi}{2l_1+1}
\frac{4\pi}{2l_2+1}\frac{k_<^{l_1}}{k_>^{l_1+1}}\frac{k_<^{l_2}}{k_>^{l_2+1}}
Y_{l_1}^{{m_1}^*}(\theta',\phi')Y_{l_1}^{m_1}(\theta,\phi)Y_{l_2}^{{m_2}^*}(\theta',\phi')Y_{l_2}^{m_2}(\theta,\phi)
\frac{\sum_{l_4m_4}\Delta_{l_4}^{m_4}(k')Y_{l_4}^{{m_4}^*}(\theta',\phi')}{\sqrt{(\hbar
vk')^2+\frac{1}{4\pi}\Delta_0^0(k')^2}}
\end{split}
\end{equation}
\end{widetext}
Here we have replaced $\sum_{\vec{k'}}$ by
$\frac{1}{(2\pi)^3}\int\,d^3\vec{k'}$ (unit volume is implicit in all expressions), and $\Delta(\vec{k'})$ in the
denominator of order parameter equation by its leading expansion
term:
$\Delta_0^0(k')Y_0^0(\theta',\phi')=\frac{1}{\sqrt{4\pi}}\Delta_0^0(k')$. We make use of the identity
\begin{widetext}
\begin{equation}\label{YYY}
\int
d\Omega'Y_{l_1}^{{m_1}^*}(\theta',\phi')Y_{l_2}^{{m_2}^*}(\theta',\phi')Y_{l_4}^{{m_4}^*}(\theta',\phi')
=\sqrt{\frac{(2l_1+1)(2l_2+1)(2l_4+1)}{4\pi}}\begin{pmatrix}
l_1 & l_2 & l_4 \\
0 & 0 & 0
\end{pmatrix}\begin{pmatrix}
l_1 & l_2 & l_4 \\
m_1 & m_2 & m_4
\end{pmatrix}
\end{equation}
\end{widetext}
where $\int d\Omega'=\int_0^{2\pi}\,d\phi'\int_0^{\pi}
\sin\theta'\,d\theta'$ and $\begin{pmatrix}
j_1 & j_2 & j_3 \\
m_1 & m_2 & m_3
\end{pmatrix}$  a Wigner 3j-symbol, which is related to the
Clebsch-Gordan coefficients. All the parameters are integers
or half-integers. Additionally, the following selection rules are
satisfied by these parameters:

1, $m_i\in\{-|j_i|,\ldots,|j_i|\}$, $i=1,2,3.$

2, $m_1+m_2+m_3=0$

3, $|j_1-j_2|\leqslant j_3\leqslant j_1+j_2$

4, $j_1+j_2+j_3$ should be an integer.

\noindent If any of the four selection rules is not satisfied, the Wigner
3j-symbol matrix is zero. Putting Eq.(\ref{YYY}) into
Eq.(\ref{a1}) gives:
\begin{widetext}
\begin{equation}\label{a2}
\begin{split}
&\sum_{l_3m_3}\Delta_{l_3}^{m_3}(k)Y_{l_3}^{{m_3}^*}(\theta,\phi)=\frac{g}{\pi}\int
k'^2\,dk'\times\\
&\sum_{l_1m_1}\sum_{l_2m_2}\sum_{l_4m_4}\sqrt{\frac{2l_4+1}{4\pi(2l_1+1)(2l_2+1)}}\begin{pmatrix}
l_1 & l_2 & l_4 \\
0 & 0 & 0
\end{pmatrix}\begin{pmatrix}
l_1 & l_2 & l_4 \\
m_1 & m_2 & m_4
\end{pmatrix}\frac{k_<^{l_1+l_2}}{k_>^{l_1+l_2+2}}
\frac{Y_{l_1}^{m_1}(\theta,\phi)Y_{l_2}^{m_2}(\theta,\phi)\Delta_{l_4}^{m_4}(k')}{\sqrt{(\hbar
vk')^2+\frac{1}{4\pi}\Delta_0^0(k')^2}}\\
\end{split}
\end{equation}
\end{widetext}
Multiplying $Y_{l_5}^{m_5}(\theta,\phi)$ and performing the integration $\int d\Omega$ we get
\begin{widetext}
\begin{equation}\label{a3}
\begin{split}
\Delta_{l_5}^{m_5}(k)&=\sum_{l_1m_1}\sum_{l_2m_2}\sum_{l_4m_4}\sqrt{(2l_4+1)(2l_5+1)}
\begin{pmatrix}
l_1 & l_2 & l_4 \\
0 & 0 & 0
\end{pmatrix}\begin{pmatrix}
l_1 & l_2 & l_4 \\
m_1 & m_2 & m_4
\end{pmatrix}\begin{pmatrix}
l_1 & l_2 & l_5 \\
0 & 0 & 0
\end{pmatrix}\begin{pmatrix}
l_1 & l_2 & l_5 \\
m_1 & m_2 & m_5
\end{pmatrix}\\
&\times\frac{g}{4\pi^2}\int_0^{\frac{\Lambda}{\hbar
v}}k'^2\,dk'\frac{k_<^{l_1+l_2}}{k_>^{l_1+l_2+2}}\frac{\Delta_{l_4}^{m_4}(k')}{\sqrt{(\hbar
vk')^2+\frac{1}{4\pi}\Delta_0^0(k')^2}}\\
\end{split}
\end{equation}
\end{widetext}
On the left side, we used the orthogonality condition of
spherical harmonics:
\begin{equation}
\int
d\Omega'Y_{l'}^{{m'}^*}(\theta,\phi)Y_l^m(\theta,\phi)=\delta_{ll'}\delta_{mm'}
\end{equation}
On the right side, we used Eq.(\ref{YYY}) to simplify the expression. We focus on the 
leading term (
$\frac{1}{\sqrt{4\pi}}\Delta_0^0(k)$ setting
$l_5=0$ and $m_5=0$) in Eq.(\ref{a3}).

According to the selection rules of Wigner 3j-symbol, we have
$l_2=l_1$, $m_2=-m_1$, $m_4=0$. The only relevant coefficient is  $\begin{pmatrix}
l_1 & l_1 & 0 \\
m_1 & -m_1 & 0
\end{pmatrix}=\frac{(-1)^{l_1-m_1}}{\sqrt{2l_1+1}}$. The gap equation is
\begin{widetext}
\begin{equation}\label{a4}
\Delta_0^0(k)=\sum_{l_1m_1,l_4}\sqrt{2l_4+1}\begin{pmatrix}
l_1 & l_1 & l_4 \\
0 & 0 & 0
\end{pmatrix}\begin{pmatrix}
l_1 & l_1 & l_4 \\
m_1 & -m_1 & 0
\end{pmatrix}\frac{(-1)^{2l_1-m_1}}{2l_1+1}\\
\frac{g}{4\pi^2}\int\frac{k_<^{2l_1}}{k_>^{2l_1+2}}\frac{\Delta_{l_4}^0(k')k'^2\,dk'}{\sqrt{(\hbar
vk')^2+\frac{1}{4\pi}\Delta_0^0(k')^2}}
\end{equation}
\end{widetext}
Note that the factor $\begin{pmatrix}
l_1 & l_1 & l_4 \\
0 & 0 & 0
\end{pmatrix}\cdot\begin{pmatrix}
l_1 & l_1 & l_4 \\
m_1 & -m_1 & 0
\end{pmatrix}$ is maximum for $l_4=0$ and decreases for $l_4\ne0$. Keeping only $l_4=0$ in $\sum_{l_4}$
in Eq.(\ref{a4}) yields
\begin{widetext}
\begin{equation}
\begin{split}
\Delta_0^0(k)&=\frac{g}{4\pi^2}\int_0^{\frac{\Lambda}{\hbar
v}}\bigg{(}\sum_{l_1}\frac{k'^2}{2l_1+1}
\frac{k_<^{2l_1}}{k_>^{2l_1+2}}\bigg{)}\frac{\Delta_0^0(k')\,dk'}{\sqrt{(\hbar
vk')^2+\frac{1}{4\pi}\Delta_0^0(k')^2}}\\
&=\frac{g}{4\pi^2}\bigg{[}\int_0^k\sum_{l_1}\frac{(k'/k)^{2l_1+2}}{2l_1+1}
\frac{\Delta_0^0(k')\,dk'}{\sqrt{(\hbar
vk')^2+\frac{1}{4\pi}\Delta_0^0(k')^2}}+\int_k^{\frac{\Lambda}{\hbar
v}}\sum_{l_1}\frac{(k/k')^{2l_1}}{2l_1+1}
\frac{\Delta_0^0(k')\,dk'}{\sqrt{(\hbar
vk')^2+\frac{1}{4\pi}\Delta_0^0(k')^2}}\bigg{]}\\
\end{split}
\end{equation}
\end{widetext}
Within this approximation the sum over $m_{1}$ can be performed as none of the terms depend on it. Since the contribution fall off as a power of $l_{1}$, the leading contribution is 
\begin{widetext}
\begin{eqnarray}
\Delta_0^0(k)=\frac{g}{4\pi^2}\int_0^k\frac{k'^2}{k^2}\frac{\Delta_0^0(k')dk'}{\sqrt{(\hbar
vk')^2+\frac{1}{4\pi}\Delta_0^0(k')^2}}
+\frac{g}{4\pi^2} \int_k^{\frac{\Lambda}{\hbar
v}}\frac{\Delta_0^0(k')dk'}{\sqrt{(\hbar
vk')^2+\frac{1}{4\pi}\Delta_0^0(k')^2}}
\end{eqnarray}
\end{widetext}
Rescaling the parameters $u=\hbar vk$ and $u'=\hbar vk'$, we
obtain the result quoted in the main text as Eq.(\ref{afinal}).
%


\end{document}